\documentclass{iopjournal}

\usepackage{graphicx}
\usepackage{amsmath}
\usepackage{subcaption}
\usepackage{xcolor}
\usepackage{mathrsfs}

\usepackage[backend=biber,style=phys]{biblatex}
\newcommand{\RomanNumeralCaps}[1]
    {\MakeUppercase{\romannumeral #1}}

\begin{document}

\articletype{Paper} 

\title{Direct experimental measurement of ion properties in extreme plasma condition}

\author{Evan M. Aguirre$^{1,2}$\textdagger} 
\author{Surabhi Jaiswal$^{3,*}$\textdagger}
\author{Sergey Khrapak$^4$}
\author{Parth Mehrotra$^3$}

\affil{$^1$University of Nevada Reno, Reno, NV 89557, United States of America}%

\affil{$^2$Department of Physics, Indian Institute of Technology Delhi, Hauz Khas, New Delhi, Delhi 110016, India}

\affil{$^3$Department of Physics, Indian Institute of Science Education and Research, Dr. Homi Bhabha Road, Pune 411008, India}%

\affil{$^4$Joint Institute for High Temperatures, Russian Academy of Sciences, 125412 Moscow, Russia}%


\affil{$^*$Author to whom any correspondence should be addressed.}

\affil{\textdagger both authors contributed equally to this study}

\email{surabhi@iiserpune.ac.in}







\begin{abstract}
We have demonstrated Laser Induced Fluorescence (LIF) in a Capacitively Coupled Plasma (CCP) discharge to directly measure the most crucial ion properties at a discharge regime suitable for a broad range of plasma research related to plasma processing and dusty plasma investigations that has been impossible for many years. The ion flow measurements in the presence and absence of dust particles show that ions move much faster directionally than expected from thermal motion, with reductions observed in the presence of dust particles. Ion temperatures are also found to exceed room temperature, contrary to a common assumption in the dusty plasma community. These findings represent a significant advancement in experimental plasma research, providing vital information to refine ion-driven process models with insights that span multiple research fields.
\end{abstract}

\section{Introduction}
\par Properties of the ion component are fundamental to many plasma applications across fields such as material processing \cite{ion_material_processing, plasma_material_processing1} space physics \cite{space_physics_ions}, and fusion energy \cite{ion_fusion_plasma}, play a crucial role in driving the physics of dusty (complex) plasma -- a plasma composed of neutrals, electrons, ions, and dust grains that is widespread in nature, from planetary rings to comet tails \cite{Goertz}, and are also important for microelectronics manufacturing \cite{dust_microeletronics}. Since ions mediate the energy, momentum and charge transfer in plasma, small variations in ion properties, such as ion temperature and ion flow, affect the macroscopic regime, process efficiency, and plasma dynamics \cite{ion_material_processing}. 
\par Plasma-based material processing typically operates from a few mTorr up to tens of mTorr in weakly ionized discharges where fractional ionization is very low ( from $\sim 10^{-6}$ to $\sim 10^{-3}$)~\cite{ccp_pressure, Lieberman2005, Fridman2008}. Such a low ionization fraction is characteristic of capacitive coupled plasma (CCP) discharge, which is one of the most common plasma processing technologies in the semiconductor industry. In CCP, these processes rely crucially on precise control of the uniformity of ion/radical fluxes and the ion angular and energy distribution function (IAEDF). Therefore, understanding ion properties under these conditions is essential for realizing plasma's ability in technological use. Despite their importance, direct, non-perturbative measurements of ion properties in technologically relevant CCP regimes remain scarce \cite{dusty_plasma_report}. 

\par Ion diagnostics are further complicated by the ubiquitous presence of charged dust particles, which form naturally in processing plasmas and are generally detrimental to stability and device yield \cite{dust_plasma_performance, Chaubey2021}. In dusty plasmas, ion flow governs dust charging, wake formation, and ion-streaming instabilities that dominate collective behavior \cite{TsytovichUFN1997,Kretschmer, melzer_crystal, ThomasNature1996,KhrapakPRL2011,jaiswal_melting_DC,Jaiswal_phase_coexistence,jaiswal_phase_bilayer, jaiswal_microgravity_ddw,SchwabeNJP2020,jaiswal_soliton,jaiswal_shock, Jaiswal_2018}. Nevertheless, ion properties in dusty plasmas have been studied primarily through theory and numerical modeling \cite{HutchinsonPRL2011,ion_wake_dustcharge, ion_wake_instability}, often assuming room-temperature ions. Direct experimental measurements of ion temperature and flow in these regimes remain exceptionally challenging \cite{Pustylnik_2021, dusty_plasma_report} and have been identified by the community as a critical unmet need \cite{dusty_plasma_report}.
\par Laser-Induced Fluorescence (LIF) provides a species-selective method for directly measuring densities, velocities, and temperature through the velocity distribution function with high temporal and spatial resolution, making it highly effective for characterizing behavior in etching and deposition plasmas~\cite{MCWILLIAMS20074860, santosh_lif_review}.  
However, LIF has been largely restricted to parameter regimes that preclude fundamental investigations of realistic plasma-processing conditions, particularly CCPs and dusty plasmas~\cite{Pustylnik_2021}. Practical limitations include weak signals in low-density or highly collisional plasmas, challenges in generating sufficiently strong species-specific fluorescence in the presence of dust, collisional quenching at pressures typical of dusty plasma experiments, and the difficulty of isolating weak ion fluorescence from background scattering~\cite{lif_review, Pustylnik_2021}.
Plasma sources with a high ionization fraction are conducive to LIF and these conditions generally occur at low pressures ($< 1$ mTorr) and high power \cite{LIF_low_pressure, Mikikian_LIF1} in accordance with Paschen’s Law. 
LIF has largely been confined to high-density, low-pressure plasmas such as inductively coupled plasmas or hot-cathode discharges~\cite{sadeghi_LIF,LIF_low_pressure,zimmerman,amy_lif_helicon,Arnas_2001_LIF}, where electron densities ($n_{e}\sim10^{13}$ cm$^{-3}$) readily support metastable excitation. In contrast, typical CCP processing plasmas operate at much lower densities, $n_{e}= (1-5)\times 10^{9}$ cm$^{-3}$~\cite{ccp_pressure}, and at pressures where collisional quenching, weak fluorescence signals, and background scattering severely limit LIF applicability.
\par Through careful experimental design and targeted improvements to the diagnostic methodology, we overcome the two primary limitations of LIF\cite{lif_review} (i) achieving a sufficient population of ions available for excitation and fluorescence, and (ii) attaining an adequate signal-to-noise ratio (SNR). This enabled us to successfully determine the ion flow velocity ($v_{i}$) and ion temperature ($T_{i}$) by measuring the ion velocity distribution function (IVDF) under discharge conditions that are directly relevant to plasma technology and where previous studies have focused on fundamental dusty plasma physics~\cite{coudel_review,ion_wake_experiment,lenaic_synchronization,lenaic_flame_melting}. Our measurements of $v_{i}$ and $T_{i}$ both in the absence and presence of dust provide critical information for refining ion-driven process models for a broad range of physics.


\section{Experimental Setup and Methods}

\par 

\begin{figure}[htb]
    \centering
    \includegraphics[width=0.9\textwidth]{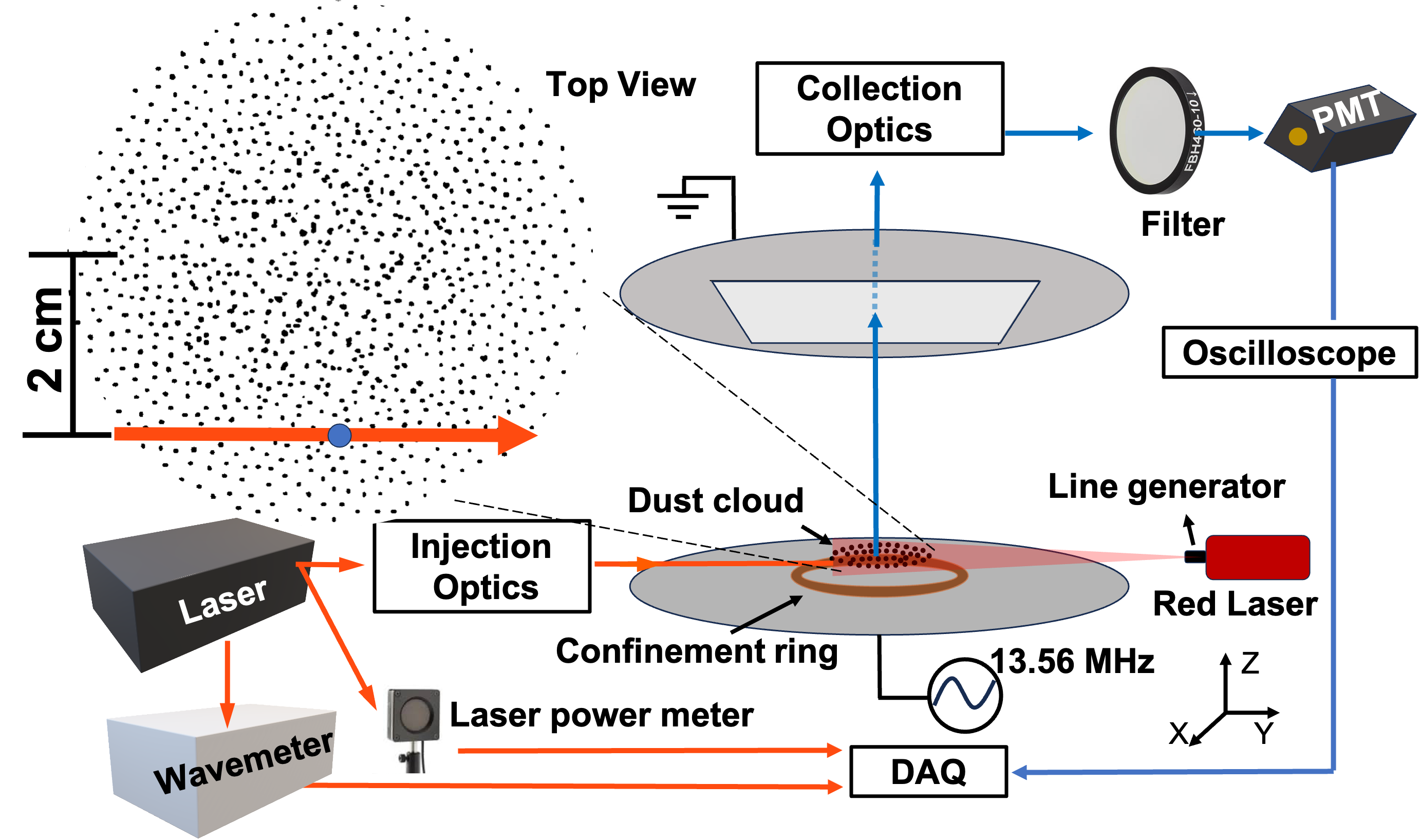}
    \caption{Schematic of the experimental arrangement. Levitated dust particles over the rf powered electrode are illuminated by a 650 nm laser. Top view shows the snapshot of the particle cloud in the plasma discharge at 10 W and $p= 8.25$ mTorr with an arrow representing the laser propagation and a blue circle representing the overlapped focus of the PMT.}
    \label{fig:schematic}
\end{figure}

\par A schematic of the experimental apparatus is shown in Fig.~\ref{fig:schematic}. The experiment was performed in a newly designed plasma reactor, where a xenon plasma was created for a range of discharge parameters. Our CCP system was manufactured with high tolerances for electrode geometry, which maximizes the power density in the plasma and ensures plasma uniformity.
The flatness tolerance of the chamber boundary, the grounded electrode and the powered electrode is all within 0.13 mm, providing exceptional parallelism between the components. Furthermore, the powered electrode has a surface roughness of no worse than 1.6 Ra $\mu$m. These specifications ensure excellent plasma confinement, which is important for ensuring metastable excitation and dust stability.

\indent Monodisperse melamine-formaldehyde spheres of radius 3.57$\pm 0.03$ $\mu$m were introduced into the plasma and formed the dust cloud levitating in the plasma sheath above the lower rf electrode. The monochrome top-view of the levitated dust particle cloud, illuminated with a 650 nm laser, is shown in Fig.~\ref{fig:schematic}. 
Inherent with any first-time measurements, there are some minor limitations to our work which should be explained here. The laser was focused at a single location in the plane of the dust which is at the sheath/plasma bulk interface. However, a change in the rf power or pressure changes the sheath such that the dust particles levitate at a slightly different height based on these two parameters. We did not calibrate our LIF measurement location to the dust layer position. Moreover, the adjustment in height from a change in plasma parameters is not larger than the laser spot size, so our measurement is representative of the ions in the plane of the dust for all discharge parameters. The LIF measurement location is 2 cm from the center axis, near the edge of the dust cloud, and all measurements were conducted in the radial ($\hat{y}$) direction (see Fig.~\ref{fig:schematic} for orientation axes).

\subsection{Laser Induced Fluorescence}

\par There are many individual advances that sum up to make a large difference in our experiment and diagnostic technique to push LIF into a more expansive parameter space compared to past efforts \cite{LIF_low_pressure, Mikikian_LIF1}, and we detail these below. Our primary advancements are: (1) use of xenon, which has low ionization and metastable levels thus creating a sufficient metastable population of ions since metastables are often directly excited from the ground state via electrons (the Xe II metastable state we use in this work is at 11.83 eV while Ar II LIF transitions used in the past ranges from 17.5 to 19 eV), (2) careful background signal subtraction, (3) recording the full PMT time series for subsequent analysis, and (4) fully de-convolving the LIF spectra with quantified broadening mechanisms. Furthermore, all glass windows of the chamber are anti-reflective in the visible range to maximize transmission of the LIF signal and excitation wavelengths.

\indent LIF is a two-step process, first absorption and then emission, to directly measure the Doppler-shifted and thermally broadened particle distribution function. For the work presented here, a laser is tuned to 605.283 nm to pump the Xe \RomanNumeralCaps{2} 5$d^{4}D_{7/2}$ metastable state to the 6$p^{4}P_{5/2}$ state, which then decays to the 6$s^{4}P_{5/2}$ state by emitting 529.369 nm photons. The laser source is a Q-switched ns pulsed Nd:YAG laser that pumps Rhodamine 5G dye at 10 Hz (Quantel QScan). A small amount of laser light is directed through a single-mode fiber to a wavemeter for wavelength monitoring (the central frequency is also known as the first moment of the laser spectral line). The absolute accuracy of our wavemeter is the primary source of our velocity error, since numerous broadening mechanisms do not affect the distribution shift and neither does significant noise. The wavemeter accuracy has an absolute accuracy for frequency measurements of 60 MHz (0.08 pm). At our laser wavelength, this corresponds to an error of $\pm 36$ m/s. The output of the laser is attenuated by multiple neutral density filters to reduce power broadening with output around 3 mJ per pulse. An energy meter measures the energy of each laser pulse for post-processing analysis. A series of mirrors guide the laser beam into the plasma chamber with a focusing lens that provides a spot size of $\sim 1-2$ mm at the measurement location. The LIF laser was injected and focused in the plane parallel to the electrode surface and is shown in Fig.~\ref{fig:schematic}. Alignment of the collection and injection optics are rigidly aligned with micrometer precision ensuring a near ideal overlap of the laser and LIF emission volume. The collection optics consist of a focusing lens, to overlap the $\sim 1-2$ mm spot size of the injected laser, a narrow bandpass filter centered on the emission wavelength, and finally a photomultiplier tube (PMT). The optics were carefully chosen to minimize laser power broadening while still maintaining a LIF signal with good SNR. Given our laser fluency of around $1\times 10^{3}$ J/m$^{2}$, we are well below the threshold for perturbations caused by high laser power on the surrounding plasma~\cite{carbone_laser}. A custom data acquisition system compiles the PMT signal, laser energy, and wavelength simultaneously and records them in files for postprocessing. 

\indent In the LIF process, we only desire the signal of fluorescence that is emitted from laser pumped atoms. To filter out any other signal we measure the background emission of 529 nm which spontaneously arises in the plasma and the scattered laser light which seeps through the bandpass filter into the PMT. We then subtract these signals from our measured PMT time series such that only the fluorescence signal remains. The ``average" steady state fluorescence, measured as a voltage signal on a fast oscilloscope (WaveRunner 6 Zi, 500 MHz) is $-3.5\times 10^{-4}$ V and the signal when the plasma is off (laser scattering) is $-1.65\times 10^{-3}$ V). As a benchmark for proper context, the fluorescence at the resonance wavelength for 10 W rf power and $p=8.25$ mTorr is $-4.5\times 10^{-3}$ V. We should note that while this difference (signal) seems small, we integrate many laser shots at each wavelength and the oscilloscope can easily measure small voltages. Past work on LIF using pulsed lasers commonly used a boxcar integrator which gates the PMT and laser pulses simultaneously and averages a signal over a set of pulses into the data acquisition program. We discard this method and instead save every PMT time series, as measured with the oscilloscope, from a laser pulse for further analysis. In this manner, we are able to change the bounds of integration of our PMT signal to maximize the SNR of the IVDF. This process produces a diagnostic technique that is more sensitive than previous efforts, which is a necessity for low signal situations and extreme plasma conditions.




\indent Xenon and our operating space present some unique challenges in obtaining the ion temperature and ion flow from the IVDF. For typical dusty plasma, it is almost universally assumed by the community that the ions are at room temperature ($\sim$ 300~K) although this has never been directly confirmed by measurements. As a result, the IVDF Doppler broadening for a heavy mass such as a xenon is expected to be roughly the same as the laser linewidth. To properly obtain the ion temperature and ion flow, the measured spectra must be deconvolved. The contribution of the deconvolution process to our overall error is much less than our main sources of error, such as the limit of the wavemeter ($\pm 36$ m/s) and our line-broadening mechanisms. The LIF emission is also spatially invariant, and the noise behavior is similar to that of white noise. Having this \textit{a priori} knowledge reduces the error inherent in the deconvolution process.

\indent The extraction of the velocity profile from LIF spectra can be done by two different methods; direct de-convolution with the help of mathematical filters or by assuming a profile and convolving it with relevant mechanisms and comparing it with the experimental data. Since very little is known about the exact nature of dust-induced noise, we used the second method. The measured LIF spectral profile is the result of the convolution of the Doppler broadening ($u_k(\nu)$), the hyperfine constants ($H(\nu)$), and the natural broadening (Lorentzian) lineshape($L(\nu)$):
\begin{equation}
    \centering
    M(\nu) = u_k(\nu) \otimes H(\nu) \otimes L(\nu)
    \label{eq1}
\end{equation}
The Doppler broadening further consists of a convolution of the velocity profile $(v(\nu))$ and the laser linewidth $(LS(\nu))$:
\begin{equation}
    \centering
    u_k(\nu) = v(\nu) \otimes LS(\nu)
    \label{eq2}
\end{equation}

The zeroth moment (total integrated intensity) is irrelevant because we are not conducting density measurements and we normalize our distributions. The first moment (central frequency) is determined by the wavemeter.  The third moment characterizes the laser profile asymmetry which in our case of a well engineered laser is not important. The laser line profile (both in space and time) is designed to be nearly 100\% Gaussian for stability. Practical values of greater than 95\% are routinely achieved. Asymmetry is only relevant for abnormal laser systems such as plasmon lattice lasers, phase-conjugate feedback (PCF) lasers, and coupled microdisk lasers. The fourth moment determines if the laser profile is more Gaussian, Lorentzian, or a combination (Voigt). This may seem fairly important since the laser profile is convolved with the measured distribution but as our de-convolution process has shown, the exact shape of the profile is not important. This arises for mainly two reasons. First, the center of the distribution is not affected by the convolution and is fixed irrespective of the laser profile's shape. In other words, the center of the distribution is the same whether or not the laser profile is purely Gaussian, purely Lorentzian, or a Voigt profile. Secondly, because we normalize the distributions, the only effect from the fourth moment would be contained in the laser linewidth which determines our temperature, but under normalization, both the Lorentzian and Gaussian widths are approximately the same and any error introduced to our calculated temperature is smaller than other experimental errors.

The laser linewidth (related to the second moment of the laser spectral line) is measured as 0.6 GHz as detailed later. Other moments of the laser line such as the zeroth moment(integrated intensity), third moment (asymmetry) and fourth moment (kurtosis) are irrelevant or ignorable. The ion temperature is determined by the width of the distribution mainly and not the wings so the exact shape (Gaussian, Lorentzian, or Voigt) of the laser profile is not important. Laser power broadening is another mechanism that must be taken into account. We do not expect much laser power broadening given our laser energy and optics design. Indeed, our measured LIF lineshape is not obviously broadened beyond recognition. However, in order to quantify this effect so our measured ion temperature is accurately resolved, we use the method detailed by Smith~\cite{smith_thesis} because we use the same Xe \RomanNumeralCaps{2} transition. All that is required is at least two measurements of the LIF spectra at different laser intensities at the same discharge parameters. For example, at the same parameters, $p=8.25$ mTorr and 10 W, we vary the laser power to measure the change in LIF signal. Because this is single photon LIF, we expect that the LIF signal should increase proportionally (i.e. linearly) with laser power. However, we notice that the LIF signal does not quite scale linearly with laser power, which means some amount of laser broadening is impacting our measurements. With two laser intensities, we determine the coefficient of power broadening $\beta$ which has units of frequency per power. Combined with our laser power $P_{L}$, we arrive at a value for the saturation of the measured LIF spectra (200 MHz for $p=8.25$ mTorr and 10 W) which manifests as an additional Lorentzian term \cite{demtroder, smith_thesis} to Eq.~(\ref{eq2}). We include this term in our deconvolution algorithm for each set of parameters to extract the proper ion temperature.


\par Assuming a Gaussian distribution first, we find a best fit Gaussian to fit the experimental data. Then, we convolve the hyperfine constant with the natural (Lorentzian) lineshape to obtain the plasma spectrum, which needs to be convolved with an assumed Gaussian Doppler broadening distribution which will give us the measured spectra. Comparison of both spectra and minimized residues is conducted to find the best-fit parameters of the Gaussian Doppler broadening. This Doppler broadening contains the true velocity distribution function, laser linewidth, and laser power broadening convolved together. With the velocity profile as a Gaussian distribution and both the laser linewidth and power broadening as Lorentzian, the Doppler broadening should essentially be a Voigt function \cite{steinberger_thesis}. The Gaussian noise for ns-pulsed dye lasers becomes significant as compared to the Lorentzian contribution. Over the course of our analysis, the difference in a Gaussian and true Voigt distribution for laser linewidth did not affect the full width half maximum value and therefore the ion temperature as under normalization both the plots become comparable. 
The distributions are parameterized using three parameters: amplitude, mean, and standard deviation. For this step, we take the same best-fit parameters obtained before, but with a reduced standard deviation because the laser linewidth is wide enough to affect the measurement and increase the measured standard deviation. The only parameter left is the true standard deviation; starting with an initial guess, we convolve it with the laser lineshape and compare it with the best-fit Doppler broadening we obtained in the previous step. With this we obtain the true standard deviation of the velocity profile and use that in the calculation of temperature. The convolution process does not change the mean of the distribution corresponding to the flow velocity in the system; hence that can be determined from the best-fit mean of the data itself. Regardless, the end result of our routine is a simulated IVDF that corresponds to the measured data and captures the true IVDF parameters.

\par The bulk flow of ions is measured via the Doppler shift from the rest wavelength given by:
\begin{equation}
    v _{i}=\frac{c\Delta f}{f_{0}+\Delta f}\approx \frac{c\Delta f}{f_{0}}=\lambda_{0}\Delta f
\end{equation}
where $v_{i}$ is the velocity of the ions, $c$ is the speed of light, $\lambda_{0}$ the rest wavelength, $f_{0}$ is the lab-frame resonant frequency, and $\Delta f$ the frequency shift ($f$ - $f_{0}$). As mentioned above, we assume for good reason that the true distribution will be Gaussian in nature. The temperature of the ions is determined by the best fit to the Gaussian distribution:
\begin{equation}
    f_{i}(v_{i})=\sqrt{\frac{m_{i}}{2\pi k_{B}T_{i}}}\text{e}^{\left(\frac{-m_{i}(v_{i}-v_{0})^{2}}{2k_{B}T_{i}}\right)}
\end{equation}
where $T_{i}$ is the temperature of the ions, $m_{i}$ is the mass of the ions, and $k_{B}$ is the Boltzmann constant.\\

\subsection{Laser Linewidth Measurement}
\par
\begin{figure}[htb]
    \centering
    \includegraphics[width=0.9\textwidth]{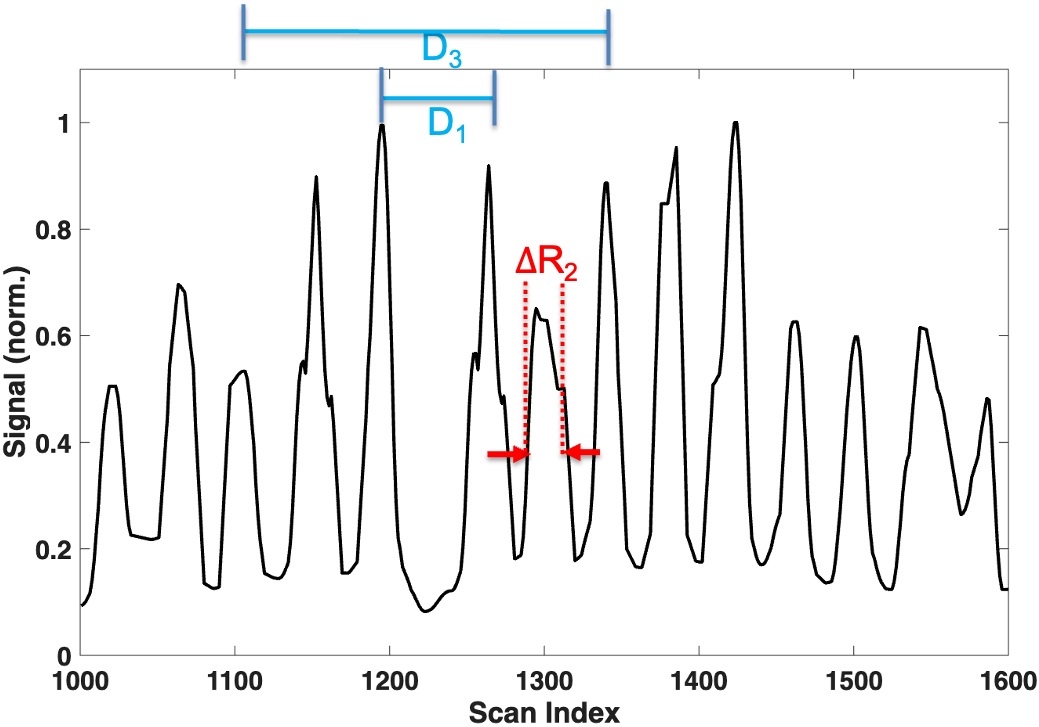}
    \caption{Interferometer pattern from the wavemeter showing the fringe quantities for calculating the laser linewidth. The scan index was limited to the central part of the pattern as the wider fringes are not necessary and the signal is normalized.}
    \label{fig:fabry}
\end{figure}

 Understanding all of the individual broadening mechanisms in LIF spectra is essential to uncover the true shape of the IVDF. In particular, the laser linewidth is a crucial parameter that greatly affects the ability to make accurate temperature measurements, especially when the Doppler broadening is of the same order. To clearly establish the laser linewidth, we have quantitatively measured it with a wavemeter's (HighFinesse WS6) interferometer. The classic method for evaluating a laser's linewidth is to use a Fabry-Perot interferometer. A comparison of different types of interferometers for measuring a ns-pulsed laser's linewidth and comparing them to the manufacturer's stated value is contained in Ref.~\cite{paul_thesis}. The fine mode of the wavemeter uses an internal solid-state Fizeau interferometer system to measure the laser linewidth. The free spectral range of the interferometer is 9 GHz which means that it can measure a maximum laser linewidth of 2.7 GHz, as the limit is about 30\% of the free spectral range. Our expected laser linewidth is much less than the limit of the system. Based on the interferometer pattern, the laser linewidth is calculated as:
\begin{equation}
    \Delta \tilde{\nu} = 4 \tilde{\nu}_{FSR} \frac{\Delta R_{2}}{(D_{3} - D_{1})}
\end{equation}
where $\tilde{\nu}_{FSR}$ is the free spectral range and the other quantities are as shown graphically in a plot of the interferometer pattern in Fig.~\ref{fig:fabry}. The laser linewidth is determined to be 0.6 GHz on the basis of Eq.~(5) and the data in Fig.~\ref{fig:fabry}.

\section{Results and Discussion}
\par
\begin{figure}
    \centering
    \includegraphics[angle=0,width=0.9\textwidth]{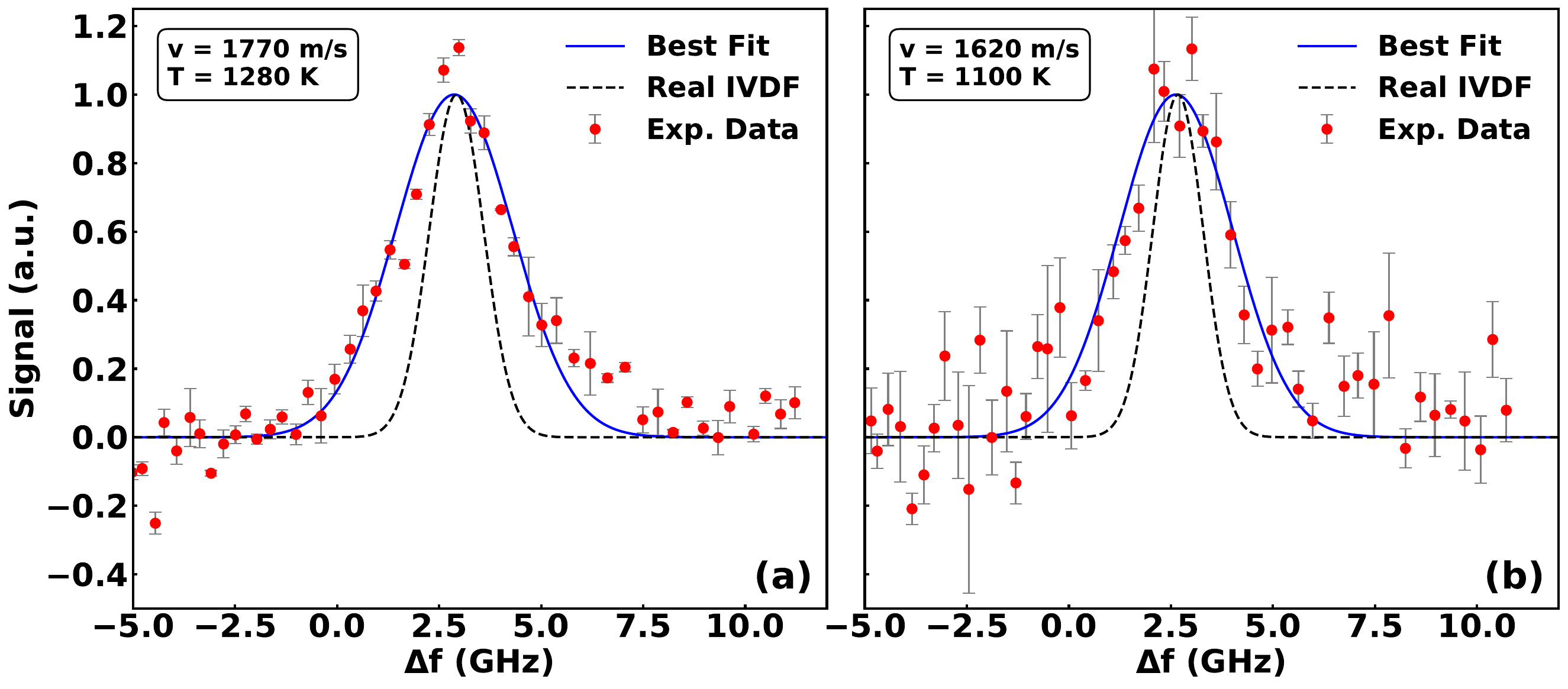}
    \caption{Ion velocity distribution functions at the sheath interface for (a) 15 W rf power and $p=6.15$ mTorr without dust and (b) 10 W rf power and $p=8.25$ mTorr with dust present in the plasma discharge. The measured IVDF was deconvoluted to get the real IVDF shown by black dashed line. }
    \label{fig:ivdf}
\end{figure}

 \indent For our lowest pressure and highest rf power that we studied, 6.15 mTorr and 15 W rf power, respectively, the IVDF has a very good SNR and is nicely fit by a Gaussian with $T_{i}\simeq1280$ K and a mean bulk flow of $v_{i}\simeq1770$ m/s, as shown in Fig.~\ref{fig:ivdf} (a). For presentation and computational purposes, the IVDF is normalized to the maximum signal. At higher pressure and lower rf power, the SNR is lower but even in the presence of dust, as shown in Fig.~\ref{fig:ivdf}(b) for a pressure of 8.25 mTorr and 10 W rf power, the real distribution of the ions yields clear values for $T_{i}\simeq1100$ K and $v_{i}\simeq1620$ m/s. The flow speed of the ions is slightly higher than the ion sound speed given by $v_{s}=\sqrt{T_{e}/m_{i}}\simeq1540$ m/s and much higher than the thermal speed of the ions given by $v_{Ti}=\sqrt{T_{i}/m_{i}}\simeq 263$ m/s (here $k_{\rm B}$ is set to unity to simplify notation and $T_{i}=1100$ K, see Fig.~\ref{fig:ivdf}(b)).  

\par
\begin{figure}[htb]
    \centering
    \includegraphics[width=0.9\textwidth]{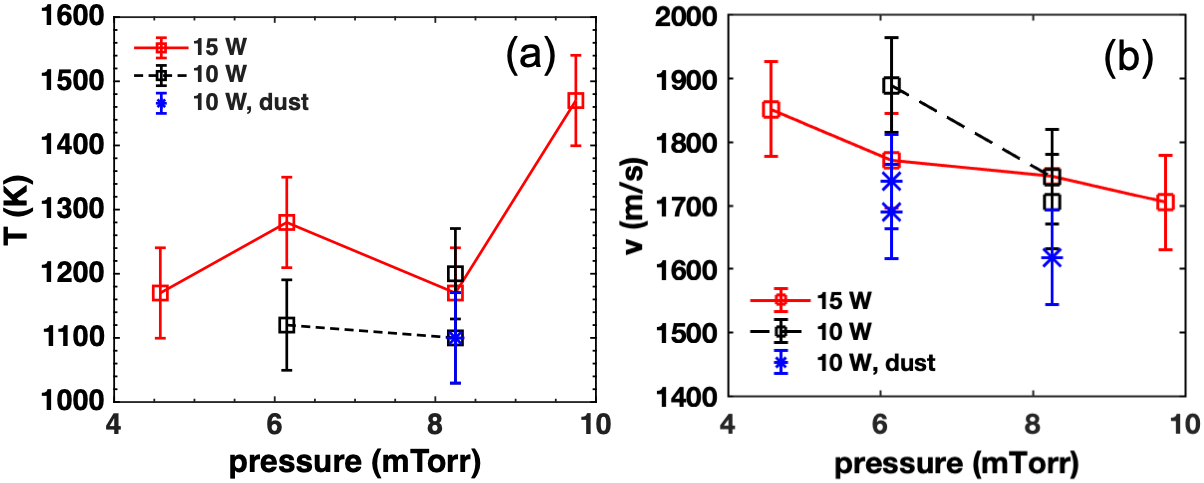}
    \caption{Ion temperature (a) and ion velocity (b) variation for different rf power levels and pressure, with and without dust. Ion temperature is nearly constant but flow velocity is reduced by over 100 m/s with the introduction of dust. }
    \label{fig:T_v_dust}
\end{figure}

 The ion temperature and the ion flow velocity, both with and without dust, as a function of rf power and pressure, are shown in Fig.~\ref{fig:T_v_dust}(a) and (b) respectively. The ion temperature is roughly the same for all the parameters, between 1100 and 1300 K, except for high rf power and pressure, the ion temperature is considerably higher at 1450 K. It is often assumed for dusty plasmas that the ions are room-temperature (300 K), but our measurement shows that there is significant heating. Notably, the ion temperature in the presence of dust is the same as when there is no dust. The flow velocity as shown in Fig.~\ref{fig:T_v_dust}(b) is gradually decreasing with increasing pressure and there is a noticeable reduction in the flow velocity in the presence of dust. For all parameters studied, $v_{i} > v_{s}$, even taking into account the measurement error. This notable finding can be useful in refining dusty plasma models for understanding the ion wake effect where the ions are affected by the presence of dust particles.

\section{Conclusion}
\par \par In conclusion, we have presented the first ever direct measurements of ion properties under previously unachievable conditions suitable for a broad range of plasma-based applications. Our measurements clearly show the importance of ions and their effect on the behavior of dust particles and vice versa. The addition of dust in the plasma reduces the ion flow velocity by over 100 m/s, which is a noteworthy finding that has not been explored experimentally. Although the background plasma parameters (e.g. $T_{e}$ and $n_{e}$) may change, thereby affecting the ion flow as well, the direct effect is that the streaming ions exchange momentum with the dust particles, and this slows down the ions. 
The measured ion temperature is above room temperature, which
should be properly accounted for in future studies. 
The measured ion flow and ion temperature provide crucial information, 
which will help to improve our current understanding and refine existing models of basic ion-dust interactions. Examples include particle charging, the ion drag force, ion streaming instability, and particle-particle interactions with particular emphasis on ion wake effects. These are responsible for a very broad range of physics observed in both gravity-based and microgravity experiments with complex plasma.
Our work also brings a trans-formative advancement to multiple fields of plasma technology by providing an accurate description of ion behavior and paves the road for future experimental, theoretical, and computational studies which will help improve the efficiency of plasma-related processes.



\ack{The authors acknowledge support from the Center for KINETIC Plasma Physics at West Virginia University. E. M. A thanks Thomas Steinberger for his assistance in acquiring data and helpful discussions. S. J. acknowledges John Earl of Eastern Michigan University for his assistance in designing the plasma chamber. }



\data{The data that support the findings of this study are available from the corresponding author upon reasonable request.}


\printbibliography

\end{document}